\documentclass[%
 reprint,
 amsmath,amssymb,
 aps
]{revtex4-1}

\usepackage[english]{babel}

\makeatletter
\def\bbl@set@language#1{%
  \edef\languagename{%
    \ifnum\escapechar=\expandafter`\string#1\@empty
    \else\string#1\@empty\fi}%
  \@ifundefined{babel@language@alias@\languagename}{}{%
    \edef\languagename{\@nameuse{babel@language@alias@\languagename}}%
  }%
  \select@language{\languagename}%
  \expandafter\ifx\csname date\languagename\endcsname\relax\else
    \if@filesw
      \protected@write\@auxout{}{\string\select@language{\languagename}}%
      \bbl@for\bbl@tempa\BabelContentsFiles{%
        \addtocontents{\bbl@tempa}{\xstring\select@language{\languagename}}}%
      \bbl@usehooks{write}{}%
    \fi
  \fi}
\newcommand{\DeclareLanguageAlias}[2]{%
  \global\@namedef{babel@language@alias@#1}{#2}%
}
\makeatother

\DeclareLanguageAlias{en}{english}

\usepackage{graphicx}
\usepackage{amsmath}
\usepackage{array}
\usepackage{amssymb}
\usepackage{dcolumn}
\usepackage{bm}
\let\savecorresponds\corresponds
\let\corresponds\relax
\usepackage{mathabx}
\usepackage{enumitem} 
\usepackage{tabularx}
\let\corresponds\savecorresponds

\def\ket#1{ | #1 \rangle}

\def\pd2v#1#2#3{\frac{\partial^2 #1}{\partial #2 \partial #3}}

\def\binom#1#2{\left( \begin{array}{c} #1 \\ #2 \end{array} \right)}

\def \2x2mat#1#2#3#4{
\left( \begin{array}{cc}
#1 &  #2 \\  #3 &  #4
\end{array} \right)
}

\begin{document}

\preprint{APS/123-QED}

\title{General model of photon-pair detection with an image sensor}

\author{Hugo Defienne}
\email{defienne@princeton.edu}
\author{Matthew Reichert}%
\author{Jason W. Fleischer}%
\affiliation{%
Department of Electrical Engineering, Princeton University, Princeton, NJ 08544, USA
}%

\date{\today}


\begin{abstract}
We develop an analytic model that relates intensity correlation measurements performed by an image sensor to the properties of photon pairs illuminating it. Experiments using both an effective single-photon counting (SPC) camera and a linear electron-multiplying charge-coupled device (EMCCD) camera confirm the model. 

\end{abstract}

\pacs{Valid PACS appear here}
\maketitle



Because it may exhibit quantum features at room temperature, light is one of the most promising platforms to investigate quantum mechanics and its applications in quantum computing, communication, and imaging~\cite{walmsley_quantum_2015}. Pairs of photons represent the simplest system showing genuine quantum entanglement in all their degrees of freedom: spatial, spectral, and polarization~\cite{brendel_pulsed_1999,kwiat_new_1995,howell_realization_2004}. Demonstrations range from fundamental tests of Bell's inequality with polarization entangled photons~\cite{aspect_experimental_1982} to the development of new imaging techniques~\cite{brida_experimental_2010}. Spatial entanglement between photons is particularly attractive, since its natural high-dimensional structure~\cite{fickler_quantum_2012,krenn_generation_2014} holds promise for powerful information processing algorithms~\cite{tasca_continuous-variable_2011,langford_measuring_2004} and secure cryptographic protocols~\cite{walborn_schemes_2008,mirhosseini_high-dimensional_2015}. While generating photon pairs entangled over a large number of spatial positions is now commonly achieved using spontaneous parametric down-conversion (SPDC)~\cite{malygin_spatiotemporal_1985}, full characterization of entangled photon states in high-dimensional Hilbert spaces remains a challenging task. Indeed, the process requires intensity correlation measurements between all pairs of possible positions, and its efficiency strongly depends on the properties of the detection system.

Light intensity correlation is a type of optical measurement used in imaging techniques, such as scintigraphy~\cite{digenis_gamma_1991} and ghost imaging~\cite{pittman_optical_1995}, and in some characterization procedures, such as dynamic light scattering~\cite{berne_dynamic_2013} and fluorescence correlation spectroscopy~\cite{magde_thermodynamic_1972}. In quantum optics, intensity correlation measurements are used to measure coincidences between correlated photons. The detection apparatus generally involves single-photon sensitive devices connected to an electronic coincidence counting circuit. The number of measurements required scales with both the number of correlated photons and the number of optical modes. Typically, correlation measurements of spatially entangled photon pairs are performed with two avalanche photo-diodes (APDs) that are raster-scanned over the different positions. Since pairs generated by a conventional SPDC source may be entangled over a very large number of spatial modes~\cite{reichert_quality_2017,reichert_massively_2017}, this raster scanning technique is prohibitively time consuming and cannot be used in practice. 

Both electron multiplying (EM)~\cite{moreau_realization_2012,edgar_imaging_2012} and intensified CCD~\cite{hamar_transverse_2010} cameras have been used to perform high-dimensional measurements. A threshold applied on the measured images allows them to operate effectively as multi-pixel single-photon counter~\cite{basden_photon_2003}. Recent works have revealed some features of entanglement between pairs of photons generated by SPDC~\cite{moreau_realization_2012,edgar_imaging_2012}, but these techniques have not retrieved the full characteristics the photon pairs, i.e., their full joint probability distribution. Moreover, the theoretical analyses associated with these works were carried out under approximations on the form of the correlation~\cite{lantz_multi-imaging_2008,tasca_optimizing_2013,lantz_optimizing_2014}. In particular, these works assumed a regime of detection in which the two photons never hit the same pixel of the camera, which is counter-intuitive when measuring pairs that are strongly correlated in position. 

In this work, we provide a general theoretical framework for intensity correlation measurements of entangled photon pairs performed with any type of detection system, with no approximation made on the source. We then compare our model to experiments performed with two different detection systems: 1) an APD-like single-photon counter (SPC) camera, implemented using an EMCCD camera with thresholding~\cite{reichert_massively_2017}, and b) a linear EMCCD camera with no threshold. Surprisingly, we show that the joint probability distribution of photon pairs can be measured using an EMCCD camera without thresholding, which provides one of the simplest techniques to characterize high-dimensional spatial entanglement of photon pairs.         
 
\begin{figure*}
\includegraphics[width=0.8 \textwidth]{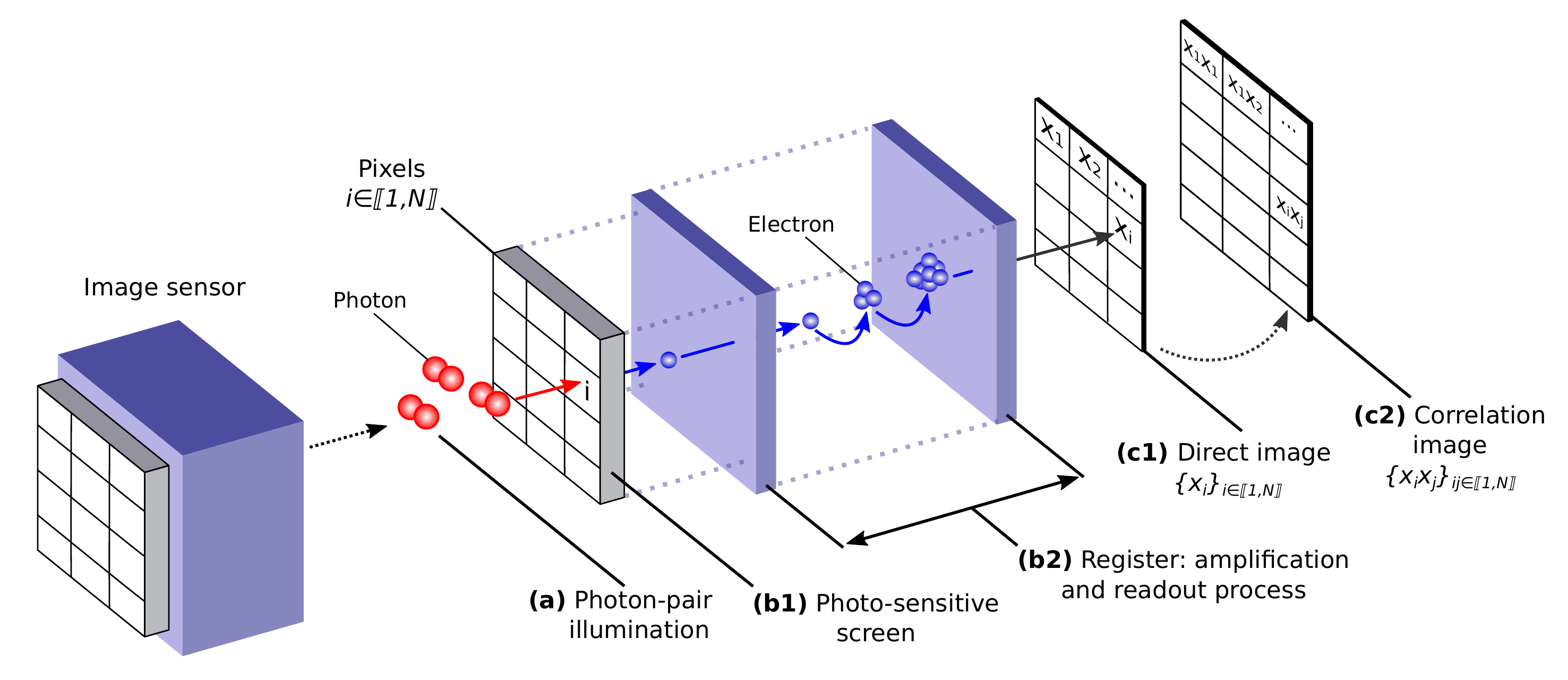}
\caption{\label{Figure1}  \textbf{Schematic of the detection architecture.} (a) A pure photon-pair state illuminates (b) the image sensor that (c) returns two types of images at the output. Distribution of photon pairs at the input is determined by the joint probability distribution $\Gamma_{ij}$, where $i$ and $j$ are two pixels of the sensor. During the exposure time photons arrive at (b1) the photo-sensitive screen and are transformed into photoelectrons with probability $\eta$. Photoelectrons in each pixel are (b2) amplified and converted into measurable signals during the readout process (c1) A direct image $ \{x_i^{(l)}\}_{i \in \ldbrack 1,N \rdbrack }$ and (c2) a correlation image $\{ x_{i}^{(l)} x_{j}^{(l)} \}_{i,j \in \ldbrack 1,N \rdbrack }$ are returned by the detector after each acquisition. }
\end{figure*}
Figure~\ref{Figure1} shows the general detection scheme considered in our model. It is studied with respect to two assumptions : 
\begin{enumerate}[label=(\roman*)]
\item Pixels of the image sensor operate independently
\item The input state is a pure two-photon state
\end{enumerate}
In the input, denoted $\ket{\phi}$, both photons have the same polarization and frequency spectrum. Its associated two-photon wavefunction depends only on the spatial properties of the pairs~\cite{smith_photon_2007} and can be expressed as
\begin{equation}
\label{etatdeuxphotons}
\ket{\phi} = \sum_{i,j \in \ldbrack 1,N \rdbrack } \phi_{ij} \ket{i,j}
\end{equation}
where $\ket{i,j}$  is a non-symmetric state defining a configuration in which the first photon of a pair is located at pixel $i$ and the second at pixel $j$, and $\phi_{ij}$ is the spatially dependent two-photon wavefunction discretized over the pixels of the sensor. The joint-probability distribution $\Gamma_{ij}=|\phi_{ij}|^2$ represents the probability of the first photon of the pair arriving at pixel $i$ and the second at pixel $j$. 

Each photon falling on the camera has a probability $\eta$ to be transformed into a photoelectron. In addition, electrons can also be generated from thermal fluctuations (dark noise). As shown in Figure~\ref{Figure1}, input electrons go through a potential amplification and readout process that converts them into detectable signals. The exact operation performed depends only on the internal characteristics of the image sensor, including the specific sensor technology and its noise properties. This process is then fully characterized by a set of conditional probability functions $\{ P(x_i|k_i) \}_{i \in \ldbrack 1,N \rdbrack }$, in which $k_i \in \mathbb{N} $ is the number of electrons present at pixel $i$ after the screen and $x_i$ the corresponding output value returned by the sensor. Henceforth, we refer to $\{ P(x_i|k_i) \}_{i \in \ldbrack 1,N \rdbrack }$ as the detector response function. 

Two types of images are returned at the output at the $l^{th}$ acquisition: a direct image, denoted $\{ x_i^{(l)} \}_{i \in \ldbrack 1,N \rdbrack }$, composed of output values returned at each pixel, and a correlation image, denoted $\{ x_{i}^{(l)} x_{j}^{(l)} \}_{i,j \in \ldbrack 1,N \rdbrack }$, computed by the tensor product of each direct image with itself. When a large number of images $M$ is recorded, averaging over all of them enables estimation of the mean values: 
\begin{align}
\langle x_i \rangle &= \lim\limits_{M \rightarrow \infty} \frac{1}{M} \sum_{l=0}^M x_i^{(l)} \label{gimeas} \\ 
\langle x_i x_j \rangle &= \lim\limits_{M \rightarrow \infty} \frac{1}{M} \sum_{l=0}^M x_i^{(l)} x_j^{(l)} \label{gigjmeas}
\end{align}
Assuming stationary illumination, $\langle x_i \rangle $ and $\langle x_i x_j \rangle $ can be written in term of their corresponding probability distributions:
\begin{align}
\langle x_i \rangle & = \sum_{x_i = 0}^{+ \infty} x_i \, P(x_j) \label{gi} \\ 
\langle x_i x_j \rangle & = \sum_{x_i = 0}^{+ \infty} \sum_{x_j = 0}^{+ \infty} x_i x_j \, P(x_i,x_j) \label{gigj}
\end{align}
where $P(x_i)$ represents the probability for the sensor to return value $x_i$ at pixel $i$ and $P(x_i,x_j)$ is the joint probability to return values $x_i$ at pixel $i$ and $x_j$ at pixel $j$, during the acquisition of each frame. Using Bayes' theorem, $\langle x_i \rangle$ can be expressed as:
\begin{equation}
\label{equ4}
\langle x_i \rangle =  \sum_{m=0}^{+ \infty} P(m) \sum_{k_i=0}^{2m} I_{k_i}  P(k_i|m) 
\end{equation}
where $P(m)$ is the probability for $m \in \mathbb{N}$ pairs to fall on the screen during the exposure time and $P(k_i|m)$ is the conditional probability of generating $k_i$ photoelectrons at pixel $i$ given $m$ pairs. $I_{k_i}$ is the mean of the detector response function at pixel $i$, defined as:
\begin{equation}
I_{k_i} = \sum_{x_i = 0}^{+ \infty} x_i \, P(x_i|k_i)
\end{equation}
$P(k_i|m)$ can then be written in terms of the marginal probability $\Gamma_i = \sum_{i} \Gamma_{ij}$ and the probability of measuring both photons of a pair at the same pixel $\Gamma_{ii}$. The complete calculation is detailed in Appendix~\ref{appeA}, and leads to the following general expression for $\langle x_i \rangle$: 
\begin{widetext}
\begin{equation}
\label{photoel1total}
\langle x_i \rangle = \sum_{m=0}^{+ \infty}  P(m) \sum_{k_i=0}^{2 m} I_{k_i}  \sum_{q=0}^{\lfloor k_i/2 \rfloor} \left( \eta^2 \Gamma_{ii}\right)^{q} \, \left( 2 \eta \, \Gamma_i-2 \eta^2 \Gamma_{ii}\right)^{k_i-2q} \left(1-2 \eta \, \Gamma_i + \eta^2 \Gamma_{ii}\right)^{m-k_i+q} {{k_i-q}\choose{q}}  \, {{m}\choose{k_i-q}}
\end{equation}
\end{widetext}
where $\binom{n}{k} = \frac{n!}{k!(n-k)!}H(n-k)$, $H$ is the Heaviside (unit step) function. Using a similar approach, the correlation image $\langle x_i x_j \rangle$ for the system to return a value $x_i$ at pixel $i$ and $x_j$ at pixel $j \neq i$ is written as:
\begin{equation}
\label{equ7}
\langle x_i x_j \rangle=\sum_{m=0}^{+ \infty } P(m) \sum _{k_i=0}^{2m} \sum_{k_j=0}^{2m} I_{k_i}I_{k_j} P(k_i,k_j |m)
\end{equation}
where $P(k_i,k_j|m)$ is the conditional probability of generating $k_i$ and $k_j$ photoelectrons at pixels $i$ and $j \neq i$, respectively, given $m$ photon pairs in each frame. Assuming that $\eta$ is uniform over the screen, the correlation coefficient $\langle x_i x_j \rangle$ can be related to the $\Gamma_{ij}$. The full calculation, given in appendix~\ref{appeB}, gives:
\begin{widetext}
\begin{align}
  \langle x_i x_j \rangle &= \sum_{m=0}^{+ \infty} P(m) \sum_{k_i=0}^{2m} \sum_{k_j=0}^{2m} I_{k_i} I_{k_j} \sum_{q=0}^{\lfloor (k_i+k_j)/2 \rfloor} \sum_{l=0}^{q} \sum_{p=0}^{q-l} (1-2 \eta \Gamma_{i}-2 \eta \Gamma_{j} +\eta^2 \Gamma_{ii} + \eta^2 \Gamma_{jj} + 2 \eta^2 \Gamma_{ij})^{m-(k_i+k_j-q)} \nonumber \\
 &\left( \eta^2 \Gamma_{jj}\right)^{p} \, \left( 2 \eta^2 \Gamma_{ij}\right)^{l} \left( \eta^2 \Gamma_{ii}\right)^{q-l-p} (2 \eta \Gamma_{i} - 2 \eta^2 \Gamma_{ii}- 2 \eta^2 \Gamma_{ij})^{k_i+l-2 (q-p)} \, (2 \eta \Gamma_{j} - 2 \eta^2 \Gamma_{jj}- 2 \eta^2 \Gamma_{ij})^{k_j-2 p-l} \nonumber \\ 
& {{k_j-l-p}\choose{p}} {{k_i-q+p}\choose{q-l-p}}  {{k_i+k_j-q-l}\choose{k_i-q+p}} {{k_i+k_j-q}\choose{l}} {{m}\choose{k_i+k_j-q}}  \label{photoel2total}
\end{align}
\end{widetext}

Equations~\ref{photoel1total} and~\ref{photoel2total} show that knowing the characteristics of the image sensor, namely its quantum efficiency and response function, as well as the number distribution of incident pairs $P(m)$, relates the direct images, $\langle x_i \rangle$ and $\langle x_j \rangle$, and the correlation image, $\langle x_i x_j \rangle$, to the joint probability distribution of the pairs $\Gamma_{ij}$. Note that these results hold only for $i \neq j$ ; the case $i=j$ is more subtle and is treated separately in appendix~\ref{appeH}. 

This set of equations provides a general link between measurements performed by any detector and the joint probability distribution of photon pairs illuminating it. We demonstrate the validity of our model by applying it to the case of an SPC camera, mimicked using a thresholded EMCCD camera~\cite{reichert_massively_2017}, and to the case of an EMCCD camera operated without threshold.

SPC cameras generally consist of an array of single-photon avalanche diodes (SPADs) or APDs with all electronics incorporated into each pixel. Photon-to-electron conversion is performed at a given quantum efficiency $\eta$, and the detector ideally returns a non-null current (value $1$) at the output if at least one electron was present at the input of the amplifier and no current if not (value $0$). As shown in appendix~\ref{appeC}, assuming a Poissonian distribution~\cite{larchuk_statistics_1995} for $P(m)$ in this model simplifies Equations~\ref{photoel1total} and~\ref{photoel2total}, allowing expression of $\Gamma_{ij}$ in terms of $\langle c_i \rangle$ and $\langle c_i c_j \rangle$:
\begin{equation}
\label{APDarray}
\Gamma_{ij}= \frac{1}{2 \eta^2 \bar{m}} \ln \left [ 1+\frac{\langle c_i c_j \rangle - \langle c_i \rangle \langle c_j \rangle}{\left(1-\langle c_i \rangle \right)\left(1-\langle c_j \rangle\right)} \right ]
\end{equation}
where the general output variable $x$ has been replaced by a binary variable $c$ (counts) that takes only two possible values $c\in\{0,1\}$. The mean photon-pair rate $\bar{m}$, which can be controlled by adjusting the exposure time of the sensor or the power of the pump laser, and the quantum efficiency $\eta$ act only as scaling factors. 

\begin{figure}[b]
\includegraphics[width=0.5\textwidth]{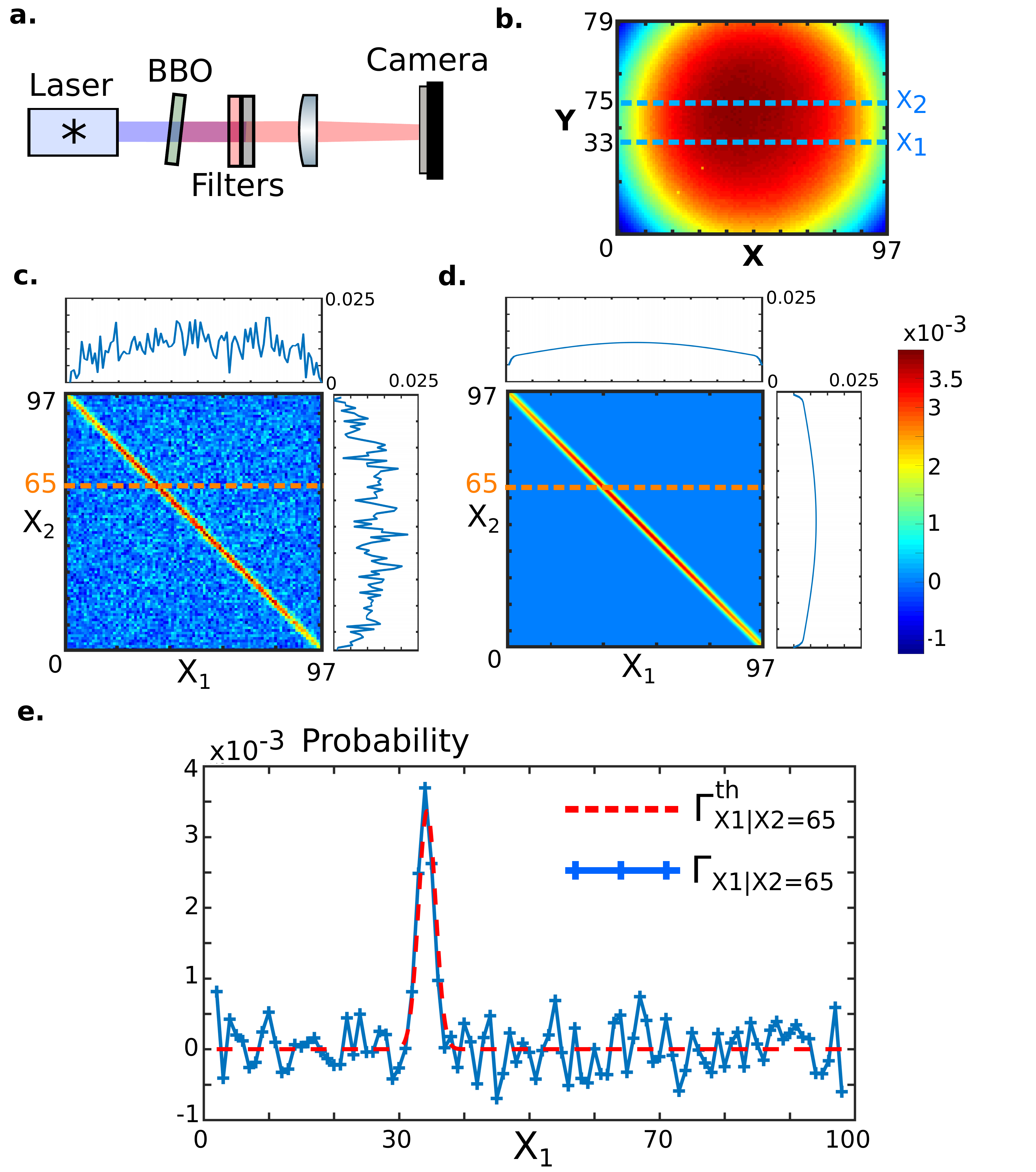}
\caption{\label{Figure2} \textbf{Measurement of the joint probability distribution of photon pairs with an SPC camera.} (a) photon pairs generated by type I SPDC are Fourier-imaged onto the screen of an EMCCD camera. A threshold applied to every image acquired at the output enables this camera to operate as an SPC camera (Appendix~\ref{appeE}). (b) Averaged direct image, proportional to the marginal distribution $\Gamma_i$. (c) 2D slices of measured joint probability distribution $\Gamma_{X_1X_2}$ at $Y_1=33$ and $Y_2=45$ [as indicated by dashed lines in (b)], and its marginals $\Gamma_{X_1}$ and $\Gamma_{X_2}$. (d) Double-Gaussian model fit $\Gamma^{th}_{X_1X_2}$ of the reconstructed joint probability distribution (Appendix~\ref{appeF}). (e) Profiles $\Gamma_{X_1|X_2=65}$ and $\Gamma^{th}_{X_1|X_2=65}$ showing the good accordance between the experiment and the double-Gaussian fit.}
\end{figure}

Experimental results are shown in Figure~\ref{Figure2}. Pairs are generated by type-I SPDC in a $\beta$-barium borate (BBO) crystal pumped by a continuous-wave (CW) laser centered at $403$ nm, and near-degenerate down-conversion is selected via spectral filters at $806 \pm 3$ nm. The far field of the output of the BBO crystal is projected onto the screen of an EMCCD camera. As detailed in~\cite{reichert_massively_2017}, applying a threshold on each acquired image effectively enables the EMCCD to operate as an SPC camera with quantum efficiency $\eta_{eff} \approx 0.44$ and a noise probability $p_{10} \approx 0.015$ (appendix~\ref{appeE}.c). To facilitate the analysis, the 4-dimensional space of pair positions is reduced to 2 dimensions by fixing the $Y$-coordinates $Y_1=33$ and $Y_2=45$ (arbitrarily) and measuring only $\langle c_i \rangle$ and $\langle c_i c_j \rangle$ along two X-axis pixels of the camera, denoted $\{X_1\}$ and $\{X_2\}$ (Figure~\ref{Figure2}.b). Figure~\ref{Figure2}.c shows the measured joint probability distribution $\Gamma_{X_1X_2}$ together with the marginals $\Gamma_{X_1}$ and $\Gamma_{X_2}$ (taken after background subtraction and normalization, appendix~\ref{appeG}). The intense anti-diagonal reveals an anti-correlated behavior of the pairs, as expected when measuring photons in the far field of the crystal. As shown in Figure~\ref{Figure2}.d, the measured joint probability distribution is well fit by a double-Gaussian model~\cite{fedorov_gaussian_2009} $\Gamma^{th}_{X_1X_2}$ of parameters $\sigma_- = 926.1 \, \mu m$ and $\sigma_+ =12.1 \, \mu m$ (appendix~\ref{appeF}). Selected profiles $\Gamma_{X_1|X_2=65}$ and $\Gamma^{th}_{X_1|X_2=65}$, highlighted in Figure~\ref{Figure2}.e, show a good match between the experiment and the double-Gaussian fit, confirming the validity of our model.

It is commonly thought that photon counting is necessary to compute the joint probability distribution of pairs of photons. We now demonstrate the surprising result that simple operation of a camera without thresholding also enables measurement of $\Gamma_{ij}$. In this case, the readout process becomes more complex, but an analytic form of $P(x|k)$ can be calculated quantitatively if the sources of noise are known, e.g. those provided in~\cite{lantz_multi-imaging_2008}. For EMCCD cameras, the mean of the detector response function $I_k$ depends linearly on the number of electrons $k$ at the input, $I_k = A \, k + x_0$, where the amplification parameter $A$ depends on the mean gain and analog-to-digital conversion and $x_0$ is a constant background (Appendix~\ref{appeE}). As shown in Appendix~\ref{appeD}, this response allows expression of $\Gamma_{ij}$ as
\begin{equation}
\label{EMCCDfull}
\Gamma_{ij} = \frac{1}{2 A^2\bar{m}\eta^2} \left[ \langle x_i x_j \rangle-\langle x_i \rangle \langle x_j \rangle \right]
\end{equation}
where the parameters $A$,  $\eta$ and $\bar{m}$ contribute only a scaling factor, which may be determined by normalization. 

\begin{figure}[b]
\includegraphics[width=0.37\textwidth]{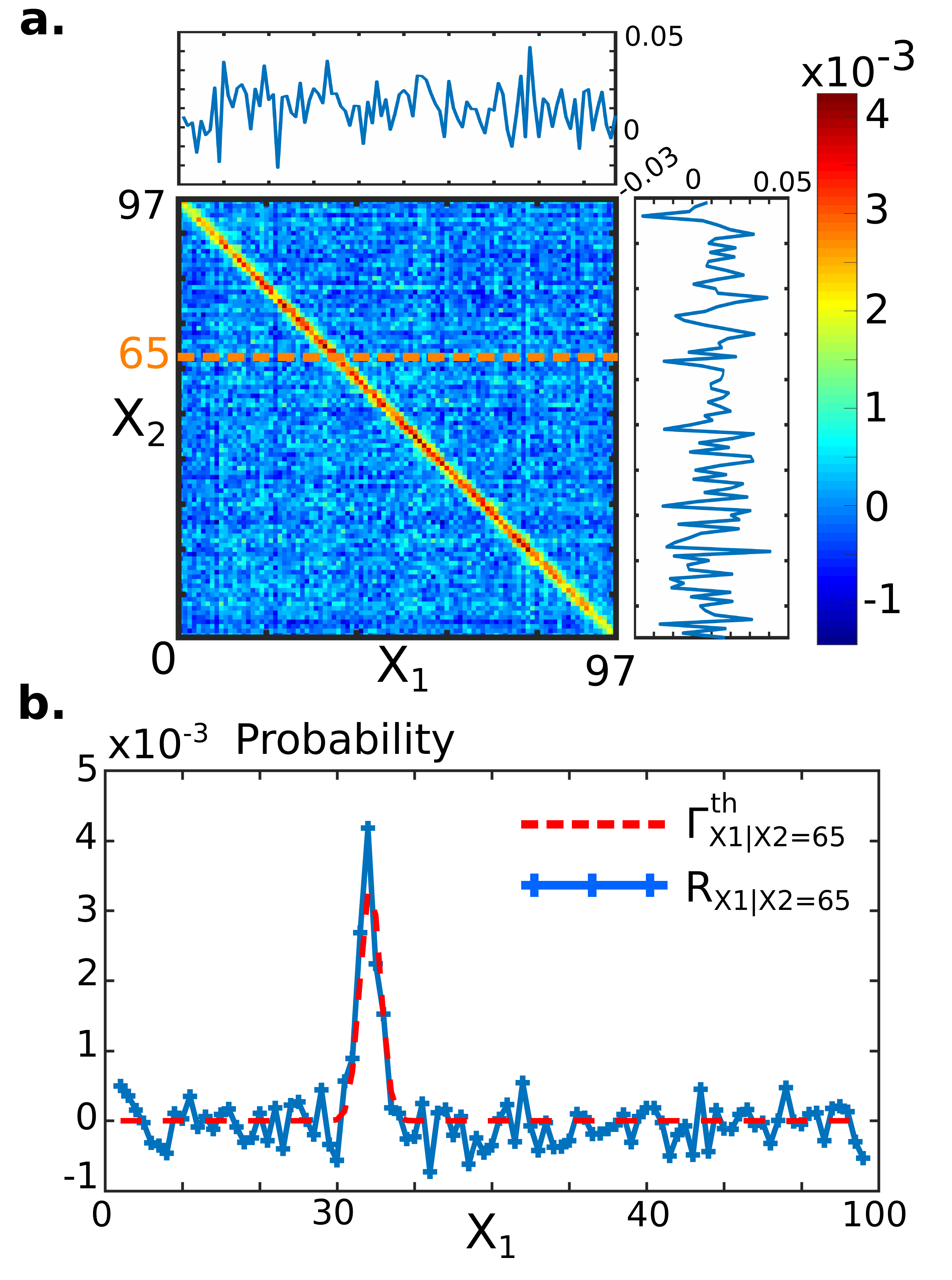}
\caption{\label{Figure3} \textbf{Measurement of the joint probability distribution of photon pairs with a non-thresholded EMCCD camera.} (a) $R_{X_1 X_2} = \langle x_{X_1} x_{X_2} \rangle - \langle x_{X_1} \rangle \langle x_{X_2} \rangle$ measured by performing an experiment in the same conditions as in Figure~\ref{Figure2}.a but without thresholding. After normalization and background subtraction, $R_{X_1 X_2}$ shows very good agreement with the theoretical model $\Gamma^{th}_{X_1 X_2}$ calculated for Figure~\ref{Figure2}.b. (b) Selected profiles $R_{X_1|X_2=65}$ (blue) and $\Gamma^{th}_{X_1|X_2=65}$ (red) confirm the good agreement.}
\end{figure}

Figure \ref{Figure3}.a shows the quantity $R_{X_1X_2} \equiv \langle x_{X_1} x_{X_2} \rangle-\langle x_{X_1} \rangle \langle x_{X_2} \rangle$ measured by performing an experiment in the same conditions as that for Figure \ref{Figure2}.a but without thresholding the output images. These results compare favorably with those of Figure~\ref{Figure2}.b. Profiles $R_{X_1|X_2=65}$ (blue) and $\Gamma^{th}_{X_1|X_2=65}$ (red) shown in Figure~\ref{Figure3}.c highlight the very good agreement between the double-Gaussian fit and the measurement without threshold.

The physical interpretation of Equation~\ref{EMCCDfull} can be seen by expanding the expression $R_{ij}$ over a finite number of images $M\gg1$ using Equations~\ref{gimeas} and~\ref{gigjmeas}:
\begin{equation}
\label{rij}
R_{ij} \approx  \frac{1}{M} \sum_{l=0}^{M} x_i^{(l)} x_j^{(l)} - \frac{1}{M^2} \sum_{l,l',l \neq l'}^{M} x_i^{(l)} x_j^{(l')} 
\end{equation}
The first term is the average tensor product of each frame with itself. Intensity correlations in this term originate from detections of both real coincidences (two photons from the same entangled pair) and accidental coincidences (two photons from two different entangled pairs). Since there is zero probability for two photons from the same entangled pair to be detected in two different images, intensity correlations in the second term originate only from photons from different entangled pairs (accidental coincidence). A subtraction between these two terms leaves only genuine coincidences, which is proportional to the joint probability distribution $\Gamma_{ij}$. 

These results show that measuring correlation between pairs of photons is not a task exclusive to single-photon sensitive devices such as SPC cameras, SPADs, or APD arrays but can be achieved using any type of image sensor. Using a megapixel image sensor as a highly parallel intensity correlator offers much promise for measuring high-dimensional entangled states, necessary for quantum computing, communication, and imaging. Moreover, the model can be extended readily to states containing more than two entangled photons, in order to study higher degrees or new forms of entanglement.

\bibliography{Biblio}

\clearpage

\appendix


\section{\label{appeA} Derivation of the general expression for $\langle x_i \rangle$ (Equations \ref{photoel1total})}

This section provides a step-by-step derivation of the general formula of $\langle x_i \rangle$ written in equation \ref{photoel1total}. Starting from the definition of $\langle x_i \rangle$ (equation~\ref{gi}) and introducing Bayes' formula gives:

\begin{align}
\langle x_i \rangle &= \sum_{x_i=0}^{+ \infty} x_i P(x_i) \nonumber \\
 &= \sum_{x_i=0}^{+ \infty} x_i  \sum_{m=0}^{+ \infty} P(m) \sum_{k_i=0}^{2 m} P(x_i|k_i) P(k_i|m) \\
&= \sum_{m=0}^{+ \infty} P(m) \sum_{k_i=0}^{2 m}  I_{k_i}P(k_i|m) \label{expbaye}
\end{align}
where $P(k_i|m)$ is the probability of generating $k_i$ photoelectrons at pixel $i$ given that a total of $m$ photon pairs reaches the sensor during the acquisition time. 

Derivation of Equation~\ref{photoel1total} relies on expressing $P(k_i|m)$ in function of the $\Gamma_{ij}$. This quantity is first expanded again using Bayes' formula as: 
\begin{equation}
\label{pkimbayes}
P(k_i|m) = \sum_{n_i=0}^{+ \infty} P(k_i|n_i) P(n_i|m) 
\end{equation}
where $P(n_i|m)$ represents the probability for $n_i$ photons to fall on pixel $i$ given $m$ photon pairs; it is calculated first in section~\ref{secpnim}. $P(k_i|m)$ is then calculated in section~\ref{pkim}. 

\subsection{\label{secpnim} Expression of $P(n_i|m)$}

We illustrate our reasoning by first calculating $P(n_i|m)$ for the case $m=1$,i.e. the case where only a single pair falls on the screen during the exposure time. Three possible events may occur:

\begin{enumerate}[label=(\alph*)]
\item Both photons reach pixel $i$: $P(2|1)=\Gamma_{ii}$
\item No photons reach pixel $i$: $P(0|1)=1-2 \Gamma_i + \Gamma_{ii}$
\item One photon of the pair reaches pixel $i$ and the other does not: $P(1|1)=2 \, (\Gamma_i- \Gamma_{ii})$
\end{enumerate}

\begin{figure}
\includegraphics[width=0.4\textwidth]{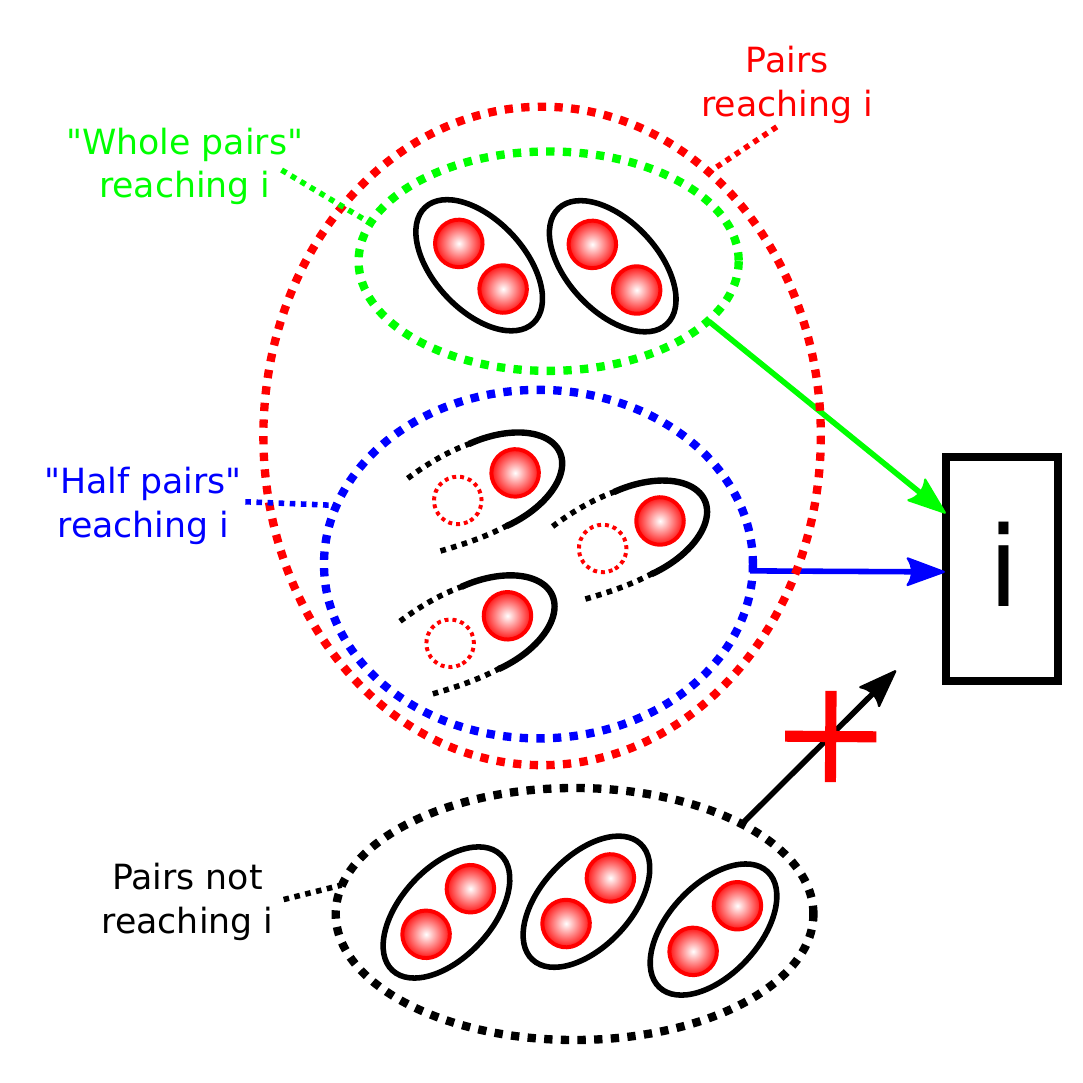}
\caption{\label{Figure5} \textbf{Types of pairs falling on pixel $i$}. Pairs falling on the screen during the exposure time can be classified in four subsets depending on their behavior relative to pixels $i$.}
\end{figure}

For $m>1$, we can simply consider each pair following one of these three possibilities. The problem is then reduced to counting all the possible configurations. As shown on Figure~\ref{Figure5}, we can label the number of pairs of each possibility:
\begin{enumerate}[label=(\alph*)]
\item $m_{2}$ is the number of pairs where both photons reach $i$ (green dashed circle)
\item $m_{1}$ is the number of pairs where only one photon reaches $i$ (blue dashed circle)
\item $m_{0}$ is the number of pairs that are lost (black dashed circle)
\end{enumerate}
As a set, the three numbers $\{m_0,m_1,m_2\}$ describe all possible arrangements of the pairs at the input. The probability for the pairs to be in a specific configuration is given by:
\begin{align}
\label{equPklkhkw}
&P_{m_0,m_1,m_2} = P(0|1)^{m_0} P(1|1)^{m_1} P(2|1)^{m_2} \nonumber \\
&{{m_1+m_2}\choose{m_1}} {{m_0+m_1+m_2}\choose{m_1+m_2}}
\end{align}
To express the probability of detecting $n_i$ photons at pixel $i$, we need to consider two conservation equations linking  $\{m_{0},m_{1},m_{2}\}$ to $\{n_i,m\}$:
\begin{align}
m&= m_{0}+m_{1}+m_{2}\\
n_i&=2 m_{2}+m_{1} 
\end{align}
With these constraints, we can write the variables $\{m_0,m_1\}$ in terms of two fixed parameters $\{n_i,m\}$ and one free parameter $m_2$:
\begin{align}
m_{1}&= n_i-2m_2 \\
m_{0}&=m-n_i+m_2
\end{align}
Since numbers are positive, $m_2 \in \ldbrack 0, \lfloor n_i/2\rfloor \rdbrack$, where $\lfloor n_i/2\rfloor$ is the integer part of  $n_i/2$. Finally, the conditional probability $P(n_i|m)$ of detecting $n_i$ photons at pixel $i$ is obtained by summing the probabilities associated with all the possible arrangements:
\begin{align}
\label{E10}
P(&n_i|m)= \sum_{m_2=0}^{\lfloor n_i/2\rfloor} \Gamma_{ii}^{m_2} (2\Gamma_i-2\Gamma_{ii})^{n_i-2 m_2} \nonumber \\
&(1-2\Gamma_i+\Gamma_{ii})^{m-n_i+m_2}{{n_i-m_2}\choose{m_2}}{{m}\choose{n_i-m_2}}
\end{align}
\subsection{\label{pkim} Expression of $P(k_i|m)$ }
Equation~\ref{pkimbayes} links $P(k_i|m)$ to $P(n_i|m)$ and $P(k_i|n_i)$. Considering the finite quantum efficiency $\eta$ of the photo-sensitive screen, $P(k_i|n_i)$ can be written as:

\begin{equation}
\label{etaeta}
P(k_i|n_i)= {{n_i}\choose{k_i}} \, \eta^{k_i} \, (1- \eta)^{n_i-k_i}
\end{equation}
Combining this expression with Equation~\ref{E10}, we find the probability for obtaining $k_i$ electrons for $m$ incident pairs:
\begin{align}
P(&k_i|m) = \sum_{n_i=0}^{2m} P(k_i|n_i)P(n_i|m) \label{demo1} \\
&= \sum_{n_i=0}^{2m} \sum_{m_2=0}^{\lfloor n_i/2 \rfloor} {{n_i}\choose{k_i}} \, \eta^{k_i} \, (1- \eta)^{n_i-k_i}\Gamma_{ii}^{m_2} \, (2 \Gamma_i-2 \Gamma_{ii})^{n_i-2m_2} \,  \nonumber \\ 
& (1-2 \Gamma_i + \Gamma_{ii})^{m-n_i+m_2} {{n_i-m_2}\choose{m_2}} \, {{m}\choose{n_i-m_2}}   \label{demo3} \\
&= \sum_{q=0}^{\lfloor k_i/2 \rfloor} \left( \eta^2 \Gamma_{ii}\right)^{q} \, \left( 2 \eta \Gamma_i-2 \eta^2 \Gamma_{ii}\right)^{k_i-2q} \, \nonumber \\
& \left( 1-2 \eta \Gamma_i + \eta^2 \Gamma_{ii}\right)^{m-k_i+q}  {{n_i-q}\choose{q}} \, {{m}\choose{n_i-q}}  \label{demo4}
\end{align}
where the transition from line~\ref{demo3} to line~\ref{demo4} follows mathematical induction on $m$. Introducing this expression in equation~\ref{expbaye} gives the final expression of $\langle x_i \rangle$ in Equation~\ref{photoel1total}.


\section{\label{appeB} Derivation of the general expression of $\langle x_i x_j \rangle$ (Equations \ref{photoel2total})}

This section provides a derivation of $\langle x_i x_j \rangle$ in Equation \ref{photoel2total}. Assuming all pixels have the same properties and are independent, only two different pixels $i$ and $j$ need to be considered. Here, we consider the case $i \neq j$ ; the case $i=j$ is treated separately in appendix~\ref{appeH}. Starting from the difinition of $\langle x_i x_j \rangle$ (equation~\ref{gigj}) and introducing Bayes'formula gives:
\begin{align}
\langle x_i x_j \rangle &= \sum_{x_i=0}^{+ \infty} \sum_{x_j=0}^{+ \infty} x_i x_j P(x_i,x_j) \nonumber \\
&=\sum_{m=0}^{+ \infty }  P(m) \sum _{k_i=0}^{2m} \sum_{k_j=0}^{2m} I_{k_i} I_{k_j} P(k_i,k_j |m)  \label{expbaye2}  
\end{align}
where $P(k_i,k_j |m)$ is the probability of generating $k_i$ photoelectrons at pixel $i$ and $k_j$ photoelectrons at pixel $j$ given that a total of $m$ photon reach the sensor during the acquisition time.
Derivation of equation~\ref{photoel2total} relies on expressing $P(k_i,k_j|m)$ as a function of $\Gamma_{ij}$. It can be expanded using Bayes' theorem as:
\begin{equation}
\label{pkikjmbayes}
P(k_i,k_j|m) = \sum_{n_i=0}^{+ \infty} \sum_{n_j=0}^{+ \infty} P(n_i,n_j|m) P(k_i|n_i) P(k_j|n_j)
\end{equation}
where: 
\begin{itemize}
\item $P(n_i,n_j|m)$ is the probability for $n_i$ and $n_j$ photons to fall on pixels $i$ and $j$, given $m$ pairs. 
\item $P(k_i|n_i) P(k_j|n_j)$ is the joint probability of generating $k_i$ and $k_j$ photoelectrons from $n_i$ and $n_j$ photons. Its factorized form relies on the assumption that pixels of the camera operate independently (no cross-talk).
\end{itemize}
An analytic form of $P(n_i,n_j|m)$ is first established in section~\ref{secpninjm}. Equation~\ref{pkikjmbayes} is then simplified to obtain an expression of $P(k_i,k_j|n_i)$ in section~\ref{pkikjm}.

\subsection{\label{secpninjm} Expression of $P(n_i,n_j|m)$}

As before, we first calculate $P(n_i,n_j|m)$ for the simple case $m=1$ and then generalize it to $m$. For $m=1$, only one pair falls on the screen during the exposure time. There are six possibilities with the following probabilities:

\begin{enumerate}[label=(\alph*)]
\item Both photons reach pixel $i$: $P(2,0|1)=\Gamma_{ii}$
\item Both photons reach pixel $j$: $P(0,2|1)=\Gamma_{jj}$
\item One photon reaches pixel $i$ and the other reaches pixel $j$: $P(1,1|1)=2\Gamma_{ij}$
\item Only one photon reaches pixel $i$, and the other does not reach $i$ or $j$ : $P(1,0|1)=2 \Gamma_{i} - 2 \Gamma_{ii}- 2 \Gamma_{ij}$
\item Only one photon reaches pixel $j$ and the other does not reach $i$ or $j$ : $P(0,1|1)=2 \Gamma_{j} - 2 \Gamma_{jj}- 2 \Gamma_{ij}$ \\
\item No photons reach either pixel $i$ or pixel $j$: $P(0,0|1)=1-2 \Gamma_{i}-2 \Gamma_{j} +\Gamma_{ii} +\Gamma_{jj} + 2 \Gamma_{ij}$. 
\end{enumerate}

\begin{figure}
\includegraphics[width=0.5 \textwidth]{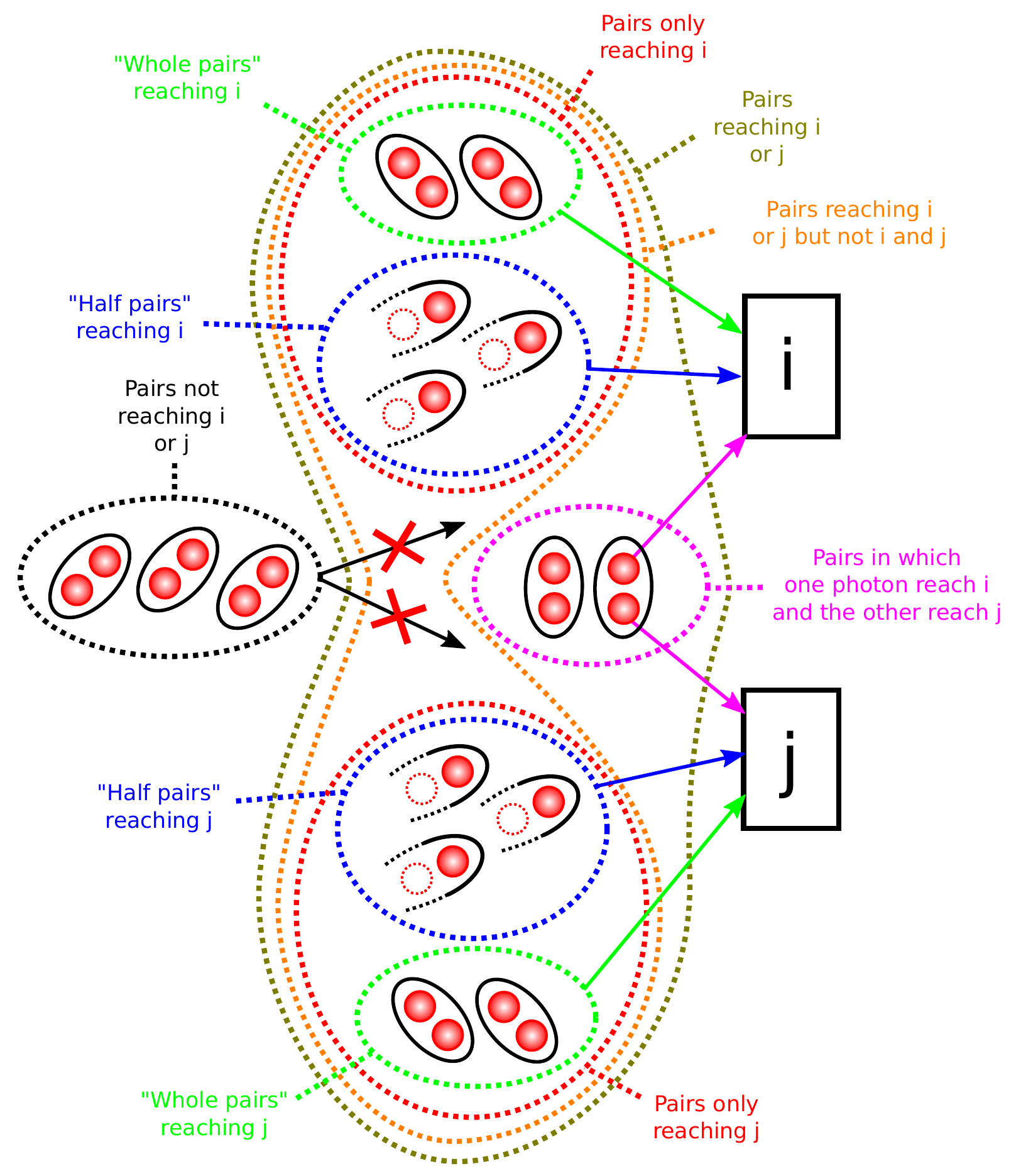}
\caption{\label{Figure6}  \textbf{Subsets of pairs falling on the screen during the exposure time relatively to pixel $i$ and $j$}. Pairs falling on the screen during the exposure time can be classified in ten subsets depending on their behavior relative to pixels $i$ and $j$. }
\end{figure}

For $m>1$, we can treat each pair as following one of these possibilities. As shown in Figure~\ref{Figure6}, the number of pairs for each possibility can be labelled as follow:
\begin{enumerate}[label=(\alph*)]
\item $m_{20}$ is the number of pairs where both photons are detected at $i$ (upper green circle)
\item $m_{02}$ is the number of pairs where both photons are detected at $j$ (lower green circle)
\item $m_{11}$ is the number of pairs where one photon is at $i$ and its pair at $j$ (pink circle)
\item $m_{10}$ is the number of pairs where one photon is at $i$ and its pair is not at $i$ or $j$  (upper blue circle)
\item $m_{01}$ is the number of pairs where one photon is at $j$ and its pair is not at $i$ or $j$ (lower blue circle) \\
\item $m_{00}$ is the number of pairs where both photons are lost (black circle)
\end{enumerate}

The six numbers $\{m_{20},m_{02},m_{11},m_{10},m_{01},m_{00}\}$ describe all possible configurations of the pairs at the input of the image sensor. The probability for the pairs to be in a specific configuration is given by:
\begin{align}
\label{Pk1Hk2h}
& P_{m_{20},m_{02},m_{11},m_{10},m_{01},m_{00}} = P(2,0|1)^{m_{20}} P(0,2|1)^{m_{02}}  \nonumber \\
& P(1,1|1)^{m_{11}} P(1,0|1)^{m_{10}} P(0,1|1)^{m_{01}} P(0,0|1)^{m_{00}} \nonumber \\
& {{m_{10}+m_{20}}\choose{m_{10}}} {{m_{01}+m_{02}}\choose{m_{01}}} \nonumber \\
& {{m_{10}+m_{01}+m_{20}+m_{02}}\choose{m_{10}+m_{01}}} \nonumber \\
& {{m_{10}+m_{01}+m_{20}+m_{02}+m_{11}}\choose{m_{11}}}\nonumber \\
& {{m_{10}+m_{01}+m_{20}+m_{02}+m_{11}+m_{00}}\choose{m_{00}}}
\end{align}
Similar to the single-pixel case, to express the probability of $n_i$ photons arriving at pixel $i$ and $n_j$ photons at pixel $j$ given a total number of pairs $m$, we need to consider three conservation equations linking $\{m_{00},m_{10},m_{01},m_{20},m_{02},m_{11} \}$ to $\{n_i,n_j,m\}$, giving the constraints:
\begin{align}
n_i&=2 m_{20}+m_{10}+m_{11} \\
n_j&=2 m_{02}+m_{01}+m_{11} \\
m&=m_{20}+m_{02}+m_{11}+m_{10}+m_{01}+m_{00}
\end{align}
Besides the three fixed parameters $\{n_i,n_j,m\}$, we also introduce one parameter $q $ defined as:
\begin{equation}
q=m_{11}+m_{20}+m_{02}
\end{equation}
$q$ represents the number of pairs in which both photons reach at least one of the two pixels. The numbers of pairs can then be expressed as written at:
\begin{align}
m_{20}&= q-m_{02}-m_{11} \\
m_{01}&= n_j-2 m_{02}-m_{11} \\
m_{10}&= n_i+m_{11}-2 (q-m_{02}) \\
m_{00}&= m-n_i-n_j+q
\end{align}
Since these numbers are positive, $\{m_{11},m_{02},q\}$ take only the following values :
\begin{align}
m_{11} &\in \ldbrack 0, q \rdbrack  \\
m_{02} &\in \ldbrack 0, q-m_{11} \rdbrack \\
q &\in \ldbrack 0, \lfloor (n_i+n_j)/2\rfloor \rdbrack 
\end{align}
The conditional probability $P(n_i,n_j|m)$ is obtained by summing the probabilities associated with all the possible configurations:
\begin{widetext}
\begin{align}
P(n_i,n_j|m)& = \sum_{q=0}^{\lfloor (n_i+n_j)/2\rfloor} \sum_{m_{11}=0}^{q} \sum_{m_{02}=0}^{q-m_{11}} (1-2 \Gamma_{i}-2 \Gamma_{j} +\Gamma_{ii} +\Gamma_{jj} + 2 \Gamma_{ij})^{m-(n_i+n_j-q)} \nonumber \\
& (2 \Gamma_{i} - 2 \Gamma_{ii}- 2 \Gamma_{ij})^{n_i+m_{11}-2 (q-m_{02})} \, (2 \Gamma_{j} - 2 \Gamma_{jj}- 2 \Gamma_{ij})^{n_j-2 m_{02}-m_{11}} \, \Gamma_{ii}^{q-m_{02}-m_{11}} \, \Gamma_{jj}^{m_{02}} \, \left( 2 \Gamma_{ij} \right)^{m_{11}} \nonumber \\
& {{n_i-q+m_{02}}\choose{q-m_{02}-m_{11}}} {{n_j-m_{11}-m_{02}}\choose{m_{02}}} {{n_i+n_j-q-m_{11}}\choose{n_i-q+m_{02}}} {{n_i+n_j-q}\choose{m_{11}}} {{m}\choose{n_i+n_j-q}} \label{pninjm}
\end{align}
\end{widetext}

\subsection{Expression of $P(k_i,k_j|m)$}
\label{pkikjm}

The probability $P(k_i,k_j|m)$ of generating $k_i$ photo-electrons at pixel $i$ and $k_j$ photo-electrons at pixel $j$ given a total of $m$ pairs is expanded starting from equation~\ref{pkikjmbayes}:

\begin{widetext}
\begin{align}
&P(k_i,k_j|m) = \sum_{n_i=0}^{2m} \sum_{n_j=0}^{2m} P(n_i,n_j|m) P(k_i|n_i) P(k_j|n_j) \label{linne1} \\
&= \sum_{n_i=0}^{2 m} \sum_{n_j=0}^{2 m} \sum_{q=0}^{\lfloor (n_i+n_j)/2\rfloor} \sum_{m_{11}=0}^{q} \sum_{m_{02}=0}^{q-m_{11}} {{n_i}\choose{k_i}}  \, \eta^{k_i} \, (1- \eta)^{n_i-k_i} {{n_j}\choose{k_j}} \, \eta^{k_j} \, (1- \eta)^{n_j-k_j} \, (1-2 \Gamma_{i}-2 \Gamma_{j} +\Gamma_{ii} +\Gamma_{jj} + 2 \Gamma_{ij})^{m-(n_i+n_j-q)} \nonumber \\
& (2 \Gamma_{i} - 2 \Gamma_{ii}- 2 \Gamma_{ij})^{n_i+m_{11}-2 (q-m_{02})} \, (2 \Gamma_{j} - 2 \Gamma_{jj}- 2 \Gamma_{ij})^{n_j-2 m_{02}-m_{11}} \, \Gamma_{ii}^{q-m_{02}-m_{11}} \, \Gamma_{jj}^{m_{02}} \, \left( 2 \Gamma_{ij} \right)^{m_{11}} \nonumber \\
& {{n_i-q+m_{02}}\choose{q-m_{02}-m_{11}}} {{n_j-m_{11}-m_{02}}\choose{m_{02}}} {{n_i+n_j-q-m_{11}}\choose{n_i-q+m_{02}}} {{n_i+n_j-q}\choose{m_{11}}} {{m}\choose{n_i+n_j-q}} \label{linne3} \\
&= \sum_{q=0}^{\lfloor (k_i+k_j)/2 \rfloor} \sum_{l=0}^{q} \sum_{p=0}^{q-l}  (1-2 \eta \Gamma_{i}-2 \eta \Gamma_{j} +\eta^2 \Gamma_{ii} + \eta^2 \Gamma_{jj} + 2 \eta^2 \Gamma_{ij})^{m-(k_i+k_j-q)} \nonumber \\
& (2 \eta \Gamma_{i} - 2 \eta^2 \Gamma_{ii}- 2 \eta^2 \Gamma_{ij})^{k_i+l-2 (q-p)} \, (2 \eta \Gamma_{j} - 2 \eta^2 \Gamma_{jj}- 2 \eta^2 \Gamma_{ij})^{k_j-2 p-l} \, \left( \eta^2 \Gamma_{ii}\right)^{q-p-l} \nonumber \\ 
& \left( \eta^2 \Gamma_{jj}\right)^{p} \, \left( 2 \eta^2 \Gamma_{ij}\right)^{l}
 {{k_i-q+p}\choose{q-p-l}} {{k_j-l-p}\choose{p}} {{k_i+k_j-q-l}\choose{k_i-q+p}} {{k_i+k_j-q}\choose{l}} {{m}\choose{k_i+k_j-q}}  \label{linne4} 
\end{align}
\end{widetext}
where the transition from line~\ref{linne3} to line~\ref{linne4} follows from mathematical induction on the variable $m$. Introducing the expression of $P(k_i,k_j|m)$ in equation~\ref{expbaye2} provides the complete expression of $\langle x_i x_j \rangle$ written in Equation~\ref{photoel2total}.


\section{ \label{appeC} Derivation of formulas linking $\Gamma_{ij}$ to $\langle x_i \rangle$ and $\langle x_i x_j \rangle$ in the case of an SPC camera (Equations~\ref{APDarray})}

The simple readout process performed in SPC camera, together with an assumption on the pair number distribution, allows simplification of Equations~\ref{photoel1total}. This operating mode is modeled by substituting the general output variable $x$ of our model by a binary variable $c$  (counts) that takes only two possible values $c\in\{0,1\}$. The corresponding conditional probability functions $P(c|k)$ are shown in Table~\ref{table1}.

\begin{table}[h]
\caption{\label{table1}%
Conditional probability functions $P(c|k)$ and mean detector response function $I_k$ that model the readout and amplification process performed by an SPC camera. $c\in \{ 0,1 \} $ is the output variable and $k$ is the number of electrons present at the input of the amplifier. $p_{10}$ is defined as the probability of generating a positive output when no photo-electrons were present at the input.}
\begin{ruledtabular}
\begin{tabular}{lcr}
\textrm{}&
$k=0$&
$k > 0$\\
\hline 
\\[-1em]
$P(c=0|k)$  & $1-p_{10}$ & $0$ \\
\\[-1em]
$P(c=1|k)$  & $p_{10}$ & $1$\\
\\[-1em]
$I_k = \sum_{c=0}^{1} c P(c|k)$
& $p_{10}$ & $1$\\
\end{tabular}
\end{ruledtabular}
\end{table}

\subsection{ \label{photoAPD} Simplification of sum over $k_j$ and $k_i$}
Using the model of readout process performed by SPC cameras (Table~\ref{table1}), $I_k$ takes the following form:
\begin{align}
I_0 &= p_{10} \\
I_{k_i} &= 1 \mbox{ if } k_i > 0
\end{align}
Using this model, Equation~\ref{photoel1total} simplifies as:
\begin{align}
&\langle c_i \rangle = \sum_{m=0}^{+ \infty}  P(m) \sum_{k_i=0}^{2 m} I_{k_i}  \sum_{q=0}^{\lfloor k_i/2 \rfloor} 
 \, \left( 2 \eta \, \Gamma_i-2 \eta^2 \Gamma_{ii}\right)^{k_i-2q} \nonumber\\  
& \left( \eta^2 \Gamma_{ii}\right)^{q} \left(1-2 \eta \, \Gamma_i + \eta^2 \Gamma_{ii}\right)^{m-k_i+q} {{k_i-q}\choose{q}}  \, {{m}\choose{k_i-q}} \label{L2} \\
&= 1-(1-p_{10}) \sum_{m=0}^{+ \infty} P(m) \left(1-2 \eta \, \Gamma_i + \eta^2 \Gamma_{ii}\right)^{m} \label{L4} 
\end{align}
and Equation~\ref{photoel2total} simplifies as:
\begin{widetext}
\begin{align}
 \langle c_i c_j \rangle &= \sum_{m=0}^{+ \infty} P(m) \sum_{k_i=0}^{2m} \sum_{k_j=0}^{2m} I_{k_i} I_{k_j} \sum_{q=0}^{\lfloor (k_i+k_j)/2 \rfloor} \sum_{l=0}^{q} \sum_{p=0}^{q-l}  (1-2 \eta \Gamma_{i}-2 \eta \Gamma_{j} +\eta^2 \Gamma_{ii} + \eta^2 \Gamma_{jj} + 2 \eta^2 \Gamma_{ij})^{m-(k_i+k_j-q)} \nonumber \\
& (2 \eta \Gamma_{i} - 2 \eta^2 \Gamma_{ii}- 2 \eta^2 \Gamma_{ij})^{k_i+l-2 (q-p)} \, (2 \eta \Gamma_{j} - 2 \eta^2 \Gamma_{jj}- 2 \eta^2 \Gamma_{ij})^{k_j-2 p-l} \, \left( \eta^2 \Gamma_{ii}\right)^{q-p-l} \nonumber \\ 
& \left( \eta^2 \Gamma_{jj}\right)^{p} \, \left( 2 \eta^2 \Gamma_{ij}\right)^{l}
 {{k_i-q+p}\choose{q-p-l}} {{k_j-l-p}\choose{p}} {{k_i+k_j-q-l}\choose{k_i-q+p}} {{k_i+k_j-q}\choose{l}} {{m}\choose{k_i+k_j-q}}  \label{summationAA}\\ 
&=1+(p_{10}^2-1-2 p_{10}) \sum_{m=0}^{+ \infty} P(m)  \left(1-2 \eta \Gamma_{i}-2 \eta \Gamma_{j} +\eta^2 \Gamma_{ii} + \eta^2 \Gamma_{jj} + 2 \eta^2 \Gamma_{ij} \right)^m \nonumber \\ 
&+p_{10} \sum_{m=0}^{+ \infty} P(m) \left[ \left(1 - \eta (2 \Gamma_i- \eta \Gamma_{ii}) \right)^m + \left(1-\eta (2 \Gamma_j- \eta \Gamma_{jj}) \right)^m \right] \label{summationBB}
\end{align}
\end{widetext}
where the transition between line~\ref{summationAA} and line~\ref{summationBB} is achieved using the following mathematical results:
\begin{itemize}
\item   $H(-l-2p)$ sets all the terms in the summation to zero except those that satisfy $l+2p=0 \Leftrightarrow l=0 \, \wedge \, p=0$
\item $H(-2(k-p)+l)$ sets all the terms in the summation to zero except those that satisfy $-2(k-p)+l=0 \Leftrightarrow l=0 \, \wedge \, k=p$
\item An extended version of the binomial theorem:
\begin{align}
&(a+b+c)^m= \nonumber \\
&\sum_{K=0}^{2m} \sum_{q=0}^{\lfloor K/2 \rfloor} a^q \, b^{K-2q} \, c^{m-K+q} {{K-q}\choose{K}} {{m}\choose{K-q}} \label{summmarante}
\end{align}
\end{itemize}
\subsection{\label{summ} Simplification of summation over $m$}

In the case of photon pairs generated through an SPDC process in a nonlinear crystal pumped by a weak continuous-wave laser, $P(m)$ can be modeled by a Poisson distribution~\cite{larchuk_statistics_1995}:
\begin{equation}
\label{poisson}
P(m) = \frac{\bar{m}^m e^{- \bar{m}}}{ m !}
\end{equation}

In this case, Equations~\ref{gi} and~\ref{gigj} simplify:
\begin{align}
\langle c_i& \rangle = 1-(1-p_{10})e^{- \bar{m} \eta \left(2 \Gamma_i- \eta \Gamma_{ii}\right)} \\
\langle c_i c_j& \rangle = 1-(1-p_{10}) \big[ e^{- \bar{m} \eta (2 \Gamma_i- \eta \Gamma_{ii})} \nonumber \\ 
&+ e^{- \bar{m} \eta \left( 2 \Gamma_i- \eta \Gamma_{jj} \right) }\big]\nonumber \\
&+ (1-p_{10})^2 e^{- \bar{m} \eta \left( 2 \Gamma_i+2 \Gamma_j - \eta \Gamma_{ii} - \eta \Gamma_{jj} - 2 \eta \Gamma_{ij} \right)  }
\end{align}
where the identity $e^a=\sum_{k=0}^{+ \infty} \frac{a^k}{k!}$ has been used. 
\subsection{Expression of $\Gamma_{ij}$ as a function of $\langle c_i \rangle $, $\langle c_j \rangle $ and $\langle c_i c_j \rangle $ (Equation~\ref{APDarray})}
Combining the two previous formulae gives:
\begin{align}
\langle c_i c_j \rangle &= -1+\langle c_i \rangle+\langle c_j \rangle \nonumber \\
&+(\langle c_i \rangle-1)(\langle c_j \rangle-1) e^{- \bar{m}  2 \eta^2 \Gamma_{ij}}  \label{gigjselongi}
\end{align}
which leads to an expression of $\Gamma_{ij}$ in terms of the direct images $\langle c_i \rangle$ and $\langle c_i \rangle$ and the correlation image $\langle c_i c_j \rangle$:
\begin{equation}
\Gamma_{ij}= \frac{1}{2 \eta^2 \bar{m}} \ln \left [ 1+\frac{\langle c_i c_j \rangle - \langle c_i \rangle \langle c_j \rangle}{(1-\langle c_i \rangle)(1-\langle c_j \rangle)} \right ]
\end{equation}

\section{\label{appeE} Model of readout process of an EMCCD camera}

Readout and amplification processes performed by an EMCCD camera  can be modeled using a quantitative model of noise described in~\cite{lantz_multi-imaging_2008}. The gray value $x$ returned by the camera at a given pixel is modeled by a random variable $X$ decomposed into

\begin{equation}
\label{valueG}
X =  \alpha \left( X^{sig}+X^{par}+X^{ser}+X^{R} \right)
\end{equation}

where:

\begin{itemize}

\item $\alpha$ is a scaling operation performed by the analog-to-digital converter.

\item $X^{sig}$ models the output value returned because of the amplification of $k$ photo-electrons generated by the photo-sensitive screen:

\begin{equation}
P_{sig}(x|k)=\frac{x^{k-1} e^{-\frac{x}{g}}}{g^k (k-1)!}
\end{equation}

where $g=(1+p_c)^L$ is the mean gain of the amplifier, where $p_c$ is the probability of duplication of an electron at each cell of the multiplication register and $L$ is the number of cells in the register. 

\item $X^{R}$ models the readout noise:

\begin{equation}
P_{R}(x)=\frac{1}{\sigma_R \sqrt{2 \pi}} e^{-\frac{(x-\mu)^2}{2 \sigma_R^2}}
\end{equation}

where $\mu$ and $\sigma_R$ define respectively the mean and standard deviation of the readout noise.

\item $X^{par}$ models the clock-induced charge (CIC) noise and the dark noise:

\begin{equation}
P_{cic}(x)=p_{par}\frac{e^{-\frac{x}{g}}}{g}
\end{equation}

where $p_{par}$ is the probability for a spurious electron to be present at the input of the multiplication register.

\item $X^{ser}$ models the electronic noise generated at each cell of the serial amplification register:

\begin{equation}
P_{ser}(x)=\sum_{l=1}^{L} p_{ser}\frac{e^{-\frac{x}{(1+p_c)^{L-l}}}}{(1+p_c)^{L-l}}
\end{equation}

where $p_{ser}$ is the probability for a spurious electron to be generated at any cell of the multiplication register.

\end{itemize}

\begin{figure}[h]
\includegraphics[width=0.5 \textwidth]{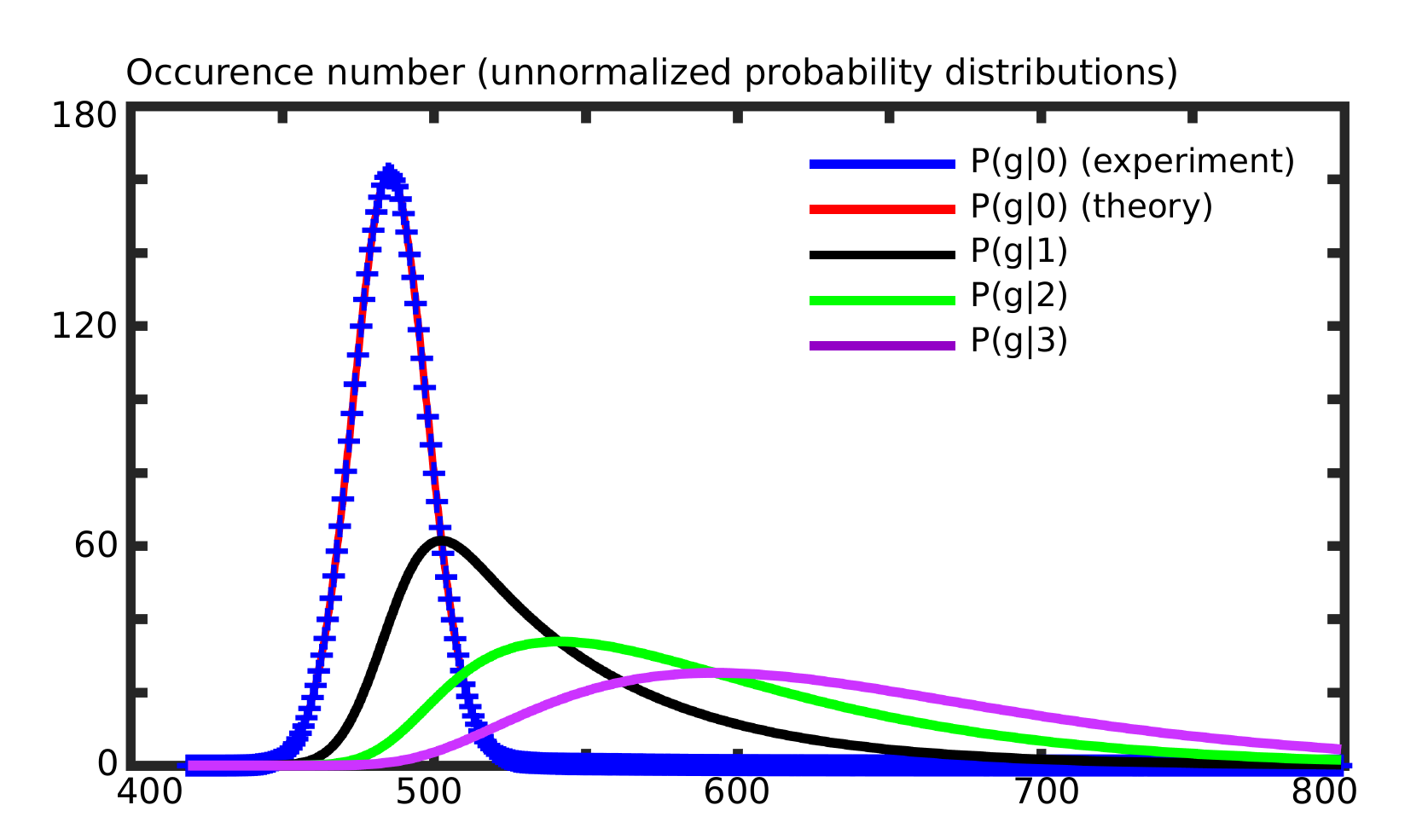}
\caption{\label{Figure7} \textbf{Model of conditional probability distributions $P(x|k)$ of an EMCCD camera.} From a initial experimental measurement of $P(x|k=0)$ performed with the shutter of the camera closed (blue), all the other conditional probability distributions $P(x|k \in \mathbb{N}^*)$ are extrapolated using the theoretical model of~\cite{lantz_multi-imaging_2008}. Only the distributions corresponding to $k \in \ldbrack 0, 3 \rdbrack $ are shown. }
\end{figure}

This model uses a total of five fitting parameters $\{\alpha , p_{ser} ,p_{par} ,\sigma_R,\mu \}$ that can be estimated from a calibration measurement $P(x|k=0)$ and the internal characteristics of the camera (in our case provided by Andor). Figure~\ref{Figure7}.a shows reconstructed response functions for $k \in \ldbrack 0, 3 \rdbrack$ as well as the calibration measurement taken for an EMCCD Andor Ixon Ultra 888 operating at a horizontal shift frequency of $10$MHz and vertical shift period of $0.6 \mu s$, at a controlled temperature of $-60^{\circ}$C. Values of the fitting parameters are: $ \{L=506, p_c = 1.37 \times 10^{-2} , \alpha = 1/19 , p_{ser} = 3.35 \times 10^{-5}  ,p_{par} = 1.23 \times 10^{-2} ,\sigma_R = 12.2 ,\mu = 25.54 \}$. 
Using this model, calculating the mean detector response function $I_{k}$ corresponds to estimating the mean value of the random variable $X$ defined by equation~\ref{valueG}:
\begin{align}
\label{approxint}
I_k =& \sum_{x=0}^{+\infty} x P_{ccd}(x|k)\\
=& \alpha \left( g k+ \mu + \, p_{par} g + \, p_{ser} \frac{g-1}{p_c} \right) \\ 
=& A \,k + x_0 \\
\end{align} 
where $A = 52.6$ and $x_0 = 569$. Linearity of the response of the EMCCD Andor iXon 888 is also confirmed by experimental tests performed by Andor.

Applying a threshold onto the images recorded at the output allows the EMCCD camera to operate as an SPC camera. Details on this particular operating mode are provided in~\cite{reichert_massively_2017}. In our experiment, the threshold is set to the value $516$ and the camera works effectively as an SPC camera with a probability $P_{10} = P(x<516|0) = 0.015$ and a quantum efficiency $\eta = 44 \%$. 



\section{ \label{appeD} Derivation of formulas linking $\Gamma_{ij}$ to $\langle x_i \rangle$ and $\langle x_i x_j \rangle$ in the case of an EMCCD camera without threshold (Equation~\ref{EMCCDfull}) }
As in equation~\ref{approxint}, we can write:
\begin{equation}
I_k = A \, k + x_0 \nonumber
\end{equation}
where $A$ is an amplification parameter and $x_0$ is a background (Details for the readout response of an EMCCD camera are given in Appendix~\ref{appeE}). This result allows simplification of Equations~\ref{photoel1total} and~\ref{photoel2total}. Derivation of Equation~\ref{EMCCDfull}, assuming a Poisson distribution for the pairs, is finally demonstrated in section~\ref{EMCCDdemo}.

\subsection{\label{photo1EMCCD} Simplification of summations over $k_i$ and $k_j$}
Simplification of $\langle x_i \rangle$ starts from Equation~\ref{photoel1total}:
\begin{widetext}
\begin{align}
\langle x_i \rangle &= \sum_{m=0}^{+ \infty}  P(m) \sum_{k_i=0}^{2 m} I_{k_i}  \sum_{q=0}^{\lfloor k_i/2 \rfloor} \left( \eta^2 \Gamma_{ii}\right)^{q} \, \left( 2 \eta \, \Gamma_i-2 \eta^2 \Gamma_{ii}\right)^{k_i-2q} \left(1-2 \eta \, \Gamma_i + \eta^2 \Gamma_{ii}\right)^{m-k_i+q} {{k_i-q}\choose{q}} \, {{m}\choose{k_i-q}}  \label{LL11} \\
&= \sum_{m=0}^{+ \infty}  P(m) \, \bigg[ x_0 \sum_{k_i=0}^{2m} \sum_{k=0}^{\lfloor k_i/2 \rfloor}  \left( \eta^2 \Gamma_{ii}\right)^{q} \, \left( 2 \eta \, \Gamma_i-2 \eta^2 \Gamma_{ii}\right)^{k_i-2q} \left(1-2 \eta \, \Gamma_i + \eta^2 \Gamma_{ii}\right)^{m-k_i+q} {{k_i-q}\choose{q}} \, {{m}\choose{k_i-q}}  \nonumber \\ 
&+ A \, \sum_{k_i=0}^{2m} k_i \sum_{k=0}^{\lfloor k_i/2 \rfloor}   \left( \eta^2 \Gamma_{ii}\right)^{q} \, \left( 2 \eta \, \Gamma_i-2 \eta^2 \Gamma_{ii}\right)^{k_i-2q} \left(1-2 \eta \, \Gamma_i + \eta^2 \Gamma_{ii}\right)^{m-k_i+q} {{k_i-q}\choose{q}} \, {{m}\choose{k_i-q}} \bigg] \label{LL22} \\
&= x_0 + 2 \,A \, \bar{m} \, \eta \, \Gamma_i \label{LL44} 
\end{align}
\end{widetext}
where $\bar{m} = \sum_{m=0}^{+ \infty} mP(m)$. The transition between line~\ref{LL11} and line~\ref{LL22} is facilitated by the identities:
\begin{equation}
\label{dervtrick}
k x^k = x \frac{d [x^k]}{dx}
\end{equation}
and
\begin{align}
\label{summmarante2}
&\sum_{k_i=0}^{2m} \sum_{k_j=0}^{2m} \sum_{k=0}^{\lfloor (k_i+k_j)/2 \rfloor} \sum_{l=0}^{k} a^{m-(k_i+k_j-k)} b^{p} \, c^l  d^{k-p-l} e^{k_i+l-2 (k-p)} \nonumber \\
&f^{k_j-2 p-l}  {{k_i-q+p}\choose{q-p-l}} {{k_j-l-p}\choose{p}} {{k_i+k_j-q-l}\choose{k_i-q+p}} \nonumber \\ 
&{{k_i+k_j-q}\choose{l}} {{m}\choose{n_i+n_j-q}}  = (a+b+c+d+e+f)^m
\end{align}

Similarly, simplification of $\langle x_i x_j \rangle$ starts from Equation~\ref{photoel2total}:

\begin{widetext}
\begin{align}
 \langle x_i x_j \rangle &= \sum_{m=0}^{+ \infty} P(m) \sum_{k_i=0}^{2m} \sum_{k_j=0}^{2m} I_{k_i} I_{k_j} \sum_{q=0}^{\lfloor (k_i+k_j)/2 \rfloor} \sum_{l=0}^{q} \sum_{p=0}^{q-l} (1-2 \eta \Gamma_{i}-2 \eta \Gamma_{j} +\eta^2 \Gamma_{ii} + \eta^2 \Gamma_{jj} + 2 \eta^2 \Gamma_{ij})^{m-(k_i+k_j-q)} \nonumber \\
 &\left( \eta^2 \Gamma_{jj}\right)^{p} \, \left( 2 \eta^2 \Gamma_{ij}\right)^{l} \left( \eta^2 \Gamma_{ii}\right)^{q-p-l} (2 \eta \Gamma_{i} - 2 \eta^2 \Gamma_{ii}- 2 \eta^2 \Gamma_{ij})^{k_i+l-2 (q-p)} \, (2 \eta \Gamma_{j} - 2 \eta^2 \Gamma_{jj}- 2 \eta^2 \Gamma_{ij})^{k_j-2 p-l} \nonumber \\ 
& {{k_i-q+p}\choose{q-p-l}} {{k_j-l-p}\choose{p}}  {{k_i+k_j-q-l}\choose{ k_i-q+p }} {{k_i+k_j-q}\choose{l}} {{m}\choose{k_i+k_j-q}} \label{summationaa1} \\
& = \sum_{m=0}^{+ \infty}  P(m) \Bigg[ x_0^2 + 2A \, x_0 \, m \, \eta \left[ \Gamma_i + \Gamma_j \right] +  A^2  \left[ 4m (m-1) \eta^2 \Gamma_i \Gamma_j + 2 m \eta^2 \Gamma_{ij} \right] \Bigg] \label{summationaa2} \\ 
& = x_0^2 + 2 A \, x_0 \, \bar{m} \eta\left[ \Gamma_i + \Gamma_j \right] +  4 A^2 (\bar{m}^2\,+\,\sigma_m^2\,-\bar{m})\eta^2\Gamma_i \Gamma_j +  2 A^2 \bar{m} \eta^2\Gamma_{ij}  \label{summationa3} 
\end{align}
\end{widetext}
where $\sigma_m^2 = \sum_{m=0}^{+ \infty} m^2 P(m)$.
\\

\subsection{\label{EMCCDdemo} Simplification of summation over $m$}

Combining Equations~\ref{LL44} and~\ref{summationa3} allows one to write
\begin{align}
\label{generalEMCCD}
\Gamma_{ij}&=\frac{1}{2 A^2 \bar{m} \eta^2} \Big[ \langle x_i x_j \rangle - \langle x_i \rangle \langle x_j \rangle \nonumber \\
&  - \frac{\sigma_m^2 - \bar{m}}{\bar{m}^2} (\langle x_i\rangle - x_0)(\langle x_j\rangle - x_0)  \Big]
\end{align}
Assuming that $P(m)$ follows a Poissonian distribution~\cite{larchuk_statistics_1995}, $\sigma_m^2 = \bar{m}$ and equation~\ref{generalEMCCD} simplifies to:
\begin{equation}
\Gamma_{ij}= \frac{1}{2 A^2\bar{m} \eta^2 } \left [ \langle x_i x_j \rangle - \langle x_i \rangle \langle x_j \rangle\right ]
\end{equation}


\section{\label{appeF} Double-Gaussian model of $\Gamma_{ij}$}

As described in~\cite{fedorov_gaussian_2009}, the joint probability distribution of photon pairs generated by SPDC (Figure~\ref{Figure2}.a) can be modeled using a double-Gaussian function of the form:
\begin{equation}
\Gamma_{ij}^{th} = a \, e^{- \frac{(x_i+x_j)^2}{4 \sigma_+^2}} e^{-\frac{(x_i-x_j)^2}{4 \sigma_-^2}}
\end{equation}
where $a$ is a normalization parameter, and $\sigma_+ =12.06\, \mu m$ and $\sigma_- = 926.12\, \mu m$ are two correlation lengths associated with the sum $(x_i+x_j)$ and difference $(x_i-x_j)$ coordinates.

\section{\label{appeG} Normalization and background removal}

If $\Gamma_{ij}$ is reconstructed directly using Equations~\ref{APDarray} or \ref{EMCCDfull}, we observe the presence of a non-zero residual background. The presence of this background can be explained by taking into account two new factors in our general model:
\begin{itemize}
\item Pump power fluctuations: intensity fluctuation of the pump implies that the assumption of a Poisson pair number distribution is no longer valid. 
\item Gain fluctuation: fluctuation of the mean gain $g$ with time (due to, e.g., temperature drifts or variations in the high voltage clock amplitude) implies that the conditional distribution $P(x|k)$ also becomes dependent on the image number $l$.
\end{itemize}

Fluctuations of $g$ induce fluctuations in both $A$ and $x_0$ over time. Equations~\ref{LL44} and~\ref{summationa3} can be generalized to take into account the mean gain fluctuations by introducing the time averaged quantities $\langle A \rangle$, $\langle x_0 \rangle$, $\langle Ax_0 \rangle$, $\langle {A^2}\rangle $ and $\langle x_0^2 \rangle$:
\begin{align}
\langle x_i \rangle &= 2 \, \langle A \rangle \,\bar{m}  \, \eta \,  \Gamma_i + \langle x_0 \rangle \\
\langle x_i x_j \rangle &=2  \langle {A^2} \rangle   \bar{m} \eta^2 \,\Gamma_{ij} + 4  \langle {A^2} \rangle  (\bar{m}^2+\sigma_{m}^2-\bar{m}) \eta^2 \,\Gamma_i \Gamma_j  \nonumber\\
&+ 2 \langle A x_0 \rangle \bar{m}\eta (\Gamma_i + \Gamma_j) +  \langle x_0^2 \rangle
\end{align}
These lead to
\begin{align}
\langle x_i x_j \rangle - \langle x_i \rangle \langle x_j \rangle &=2\langle A^2 \rangle \bar{m} \eta^2 \, \Gamma_{ij} \nonumber \\
&+4   \left[  \langle A^2 \rangle(\bar{m}^2+\sigma_m^2-\bar{m})- \langle A  \rangle^2\bar{m}^2 \right] \eta^2\Gamma_i \Gamma_j \nonumber \\
&+2  \left[\langle A x_0 \rangle-\langle A \rangle \langle x_0 \rangle \right] \bar{m} \eta^2  (\Gamma_i + \Gamma_j) \nonumber \\
&+ \langle x_0^2 \rangle -\langle x_0 \rangle^2 \\
&= 2\langle A^2 \rangle \bar{m}  \eta^2 \Gamma_{ij}+ B(\Gamma_i,\Gamma_j)
\end{align}
where $B(\Gamma_i,\Gamma_j)$ is the residual background. It can be mitigated in two ways:
\begin{enumerate}
\item First estimate $\langle x_i \rangle \langle x_j \rangle$ using only successive frames, rather than from the sum of all of them:
\begin{equation}
\langle x_i \rangle \langle x_j \rangle \approx \lim\limits_{M \rightarrow +\infty} \frac{1}{(M-1)^2}\sum_{l=1}^{M-1} x_i^{(l)} x_j^{(l+1)}
\end{equation}
The use of successive frames decreases values of the (co)variances $\langle A^2 \rangle -\langle A \rangle^2$, $\langle A x_0 \rangle-\langle A \rangle\langle x_0 \rangle$ and $\langle x_0^2 \rangle -\langle x_0 \rangle^2$, particularly when the fluctuations are relatively slowly varying.
\item Assume that $\Gamma_{ij}$ has a higher spatial frequency spectrum than $\Gamma_i$ and apply a low-pass filter on the reconstructed image to filter out the term $B(\Gamma_i,\Gamma_j)$.
\end{enumerate}

A similar procedure is also used to process reconstructed data in the SPC case, as discussed in~\cite{reichert_massively_2017}. Finally, the reconstructed $\Gamma_{ij}$ is normalized to ensure that $\sum_{ij} \Gamma_{ij} = 1$.\\

\section{\label{appeH} Case $i=j$}

In the case $i=j$, $\Gamma_{ij}=\Gamma_{ii}$ is expressed as a function of $\langle x_i^2 \rangle$ and $\langle x_i \rangle$. Similar to the calculation of $\langle x_i \rangle$,  $\langle x_i^2 \rangle$ can be written as:
\begin{widetext}
\begin{align}
\langle x_i^2 \rangle &= \sum_{i=0}^{+ \infty} x_i^2 P(x_i) \\
&=\sum_{m=0}^{+ \infty } P(m) \sum _{k_i=0}^{2m} J_{k_i}  P(k_i|m)\\
&= \sum_{m=0}^{+ \infty}  P(m) \sum_{k_i=0}^{2 m} J_{k_i}  \sum_{q=0}^{\lfloor k_i/2 \rfloor} \left( \eta^2 \Gamma_{ii}\right)^{q} \, \left( 2 \eta \, \Gamma_i-2 \eta^2 \Gamma_{ii}\right)^{k_i-2q} \left(1-2 \eta \, \Gamma_i + \eta^2 \Gamma_{ii}\right)^{m-k_i+q} {{k_i-q}\choose{q}}  \, {{m}\choose{k_i-k}}
\end{align}
\end{widetext}
where:
\begin{equation}
J_{k_i} = \sum_{x_i=0}^{+ \infty} x_i^2 P(x_i|k_i)
\end{equation}

In the case of an SPC camera, the form of $P(c|k)$ (Table~\ref{table1}) shows that $J_k = I_{k}$. Indeed, $\langle c_i^2  \rangle = \langle c_i  \rangle$ and it is not possible to reconstruct $\Gamma_{ii}$. 

In the case of an EMCCD camera, the model of the response function provided in~\cite{lantz_multi-imaging_2008} shows that $J_{k}$ is quadratic in $k$:
\begin{align}
 \sum_{x=0}^{+\infty} x^2 P(x|k) &= \alpha^2 \left( g^2 k+ \sigma_R^2 + \, p_{par} g^2 + \, p_{ser} \frac{g^2-1}{p_c(p_c+2)} \right) \nonumber \\
 &+I_k^2  \nonumber \\
 &=  [A k + x_0]^2 + A^2 k + \sigma_0^2
\end{align} 
where $\sigma_0^2$ is the variance of the background. $\langle x_i^2 \rangle$ can be written as
\begin{align}
\langle x_i^2 \rangle &= 2 A^2 \bar{m} \eta^2 \Gamma_{ii} + 4 A^2 (\bar{m} + \sigma_m^2 -\bar{m}) \eta^2 \Gamma_i^2 \nonumber \\
&+ 4  (A^2+Ax_0) \bar{m}\eta\Gamma_i + \sigma_0^2+x_0^2
\end{align}
which can be solved for $\Gamma_{ii}$.
\bibliography{Biblio}

\end{document}